\documentstyle[aps,prl,preprint,epsfig]{revtex}
\def\ts{\thinspace}
\begin{document}
\draft
\title{Tevatron Constraints on Topcolor-Assisted Technicolor}
\author{Yumian Su,\cite{YS} Gian Franco Bonini,\cite{GF} and Kenneth 
Lane\cite{KL}}
\address{Department of Physics, Boston University, Boston, Massachusetts, 
 02215}
\date{\today}
\maketitle
\widetext
\begin{abstract}
We study the constraints on models of topcolor-assisted technicolor
arising from measurements of high-$E_T$ jets and high-mass lepton pairs
at the Tevatron collider. Existing data can eliminate models that have
appeared in the literature.
\end{abstract}
\narrowtext

\paragraph*{}
The dynamical electroweak symmetry breaking scheme known as
topcolor-assisted technicolor (TC2) was proposed~\cite{Hill} to resolve the
difficulties of top-condensate models~\cite{topcondref,topcref} and of
technicolor models~\cite{tcref,etc}. Technicolor, augmented by extended
technicolor, has been unable to explain the large mass of the top quark
without conflict with precision electroweak measurements~\cite{zbbth} or
unacceptable fine tuning of parameters. On the other hand, models with
electroweak symmetry breaking induced by strong topcolor interactions that
are consistent with precision electroweak measurements require a topcolor
energy scale much higher than the electroweak scale and, therefore, very
severe fine tuning. These problems are ameliorated in TC2.
Technicolor interactions at the electroweak scale are responsible for
electroweak symmetry breaking, and extended technicolor generates the hard 
masses of all quarks and leptons {\it except} that of the top quark. Strong
topcolor interactions, broken near 1~TeV, induce a large top condensate
and all but a few GeV of the top mass, but contribute little to electroweak
symmetry breaking.

\paragraph*{}
In the simpler TC2 models, an (as yet unspecified) extended technicolor
gauge group breaks down to\cite{Hill,Lane_Eichten,Lane,lane:japan}:
\begin{equation}
   G_{TC}  \otimes  SU(3)_1  \otimes  U(1)_1  \otimes  SU(3)_2 
                \otimes  U(1)_2   \otimes  SU(2)_L 
\end{equation} 
with coupling constants $g_{TC}$, $g_1$, $g_1^\prime$, $g_2$, $g_2^\prime$,
and  $g$, respectively, where  $g^2_1\gg g^2_2$ and $g_1^{\prime 2}\gg
g_2^{\prime 2}$. At an energy scale of order 1~TeV, the gauge symmetries in
Eq.~(1) break down into the usual color and hypercharge groups:
$SU(3)_1\otimes  SU(3)_2 \rightarrow SU(3)_c$, $U(1)_1\otimes  U(1)_2
\rightarrow U(1)_Y$. Because of this breaking, the model has eight
color-octet vector bosons and one neutral $Z'$, all of which have masses of 
order $1 \mbox{ TeV}$. In the models of Ref.\cite{Lane}, $SU(3)_1$ couples
only to third-generation quarks, but the strong $U(1)_1$ couples to {\em
all} fermions. This was necessary to achieve mixing between the third
generation of quarks and the two lighter generations while ensuring,
among other things, that gauge anomalies cancelled. Thus, only the $Z'$
couples strongly to the first two families of quarks and leptons. In what
follows, we show that Tevatron experiments on high-$E_T$ jets and
high-invariant-mass lepton pairs already put stringent constraints on the
$Z'$ mass and couplings for models such as those in
Refs.~\cite{Lane,lane:japan}.

\paragraph*{}
At the Tevatron energy  
the $Z'$  interactions relevant for jet production are well approximated by
four-fermion terms:\footnote{Here and throughout, we ignore the
energy-dependent width of the $Z'$.}
\widetext
\begin{equation}
{\cal L}_{qq} = - \frac{g_{Z^\prime}^2}{2M_{Z^\prime}^2}
      \left[\sum\limits_{q=u,d,c,s}(b\ts \bar{q}_L\gamma_\mu q_L
                            +b^\prime \ts \bar{q}_R\gamma_\mu q_R)
 + d(\bar{t}_L\gamma_\mu t_L + \bar{b}_L\gamma_\mu b_L)
            + d'\ts \bar{t}_R\gamma_\mu t_R
            + d''\ts \bar{b}_R\gamma_\mu b_R\right]^2.
\end{equation}
where  $g_{Z^\prime}=\sqrt{{g_1^\prime}^2+{g_2^\prime}^2}$.
The $Z'$ interaction modifying $\ell^+\ell^-$ production is
\begin{equation}
{\cal L}_{q\ell} = -\frac{g_{Z^\prime}^2}{M_{Z^\prime}^2}
        \sum\limits_{q=u,d,c,s}(b\bar{q}_L\gamma^\mu q_L
                             +b^\prime\bar{q}_R\gamma^\mu q_R)
 \sum\limits_{\ell=e,\mu}(a\bar{\ell}_L\gamma_\mu \ell_L
                              +a\bar{\ell}_R\gamma_\mu \ell_R).
\end{equation}

\narrowtext

\noindent The $U(1)_1$ hypercharges $a$, $b$, $b'$, $d$, $d'$, $d''$ are
expected to be not much less than one. The charge $a$ must be the same for
left and right-handed electrons to avoid large atomic parity
violation when $M_{Z'} \sim 1$~TeV~\cite{lane:japan}. In the computations
that follow, we have taken $d=b$ and $d'=d''=b'$ for the $b$-quark terms,
and have ignored the $t$-quark terms. This simplification has negligible
effect on our results.

\paragraph*{}
We start by considering dijet production. The relevant
leading-order parton-level cross sections that are modified by $Z'$ exchange 
are:
\widetext
\begin{equation}
\begin{array}{lll}
&\frac{d\hat{\sigma}}{d\hat{t}}(qq\to qq) =& \frac{4\pi}{9\hat 
s^2}\alpha_s^2\left[\frac{\hat{s}^2
+\hat{u}^2}{\hat{t}^2}+\frac{\hat{s}^2+\hat{t}^2}{\hat{u}^2}
    -\frac{2}{3}\frac{\hat{s}^2}{\hat{t}\hat{u}}\right]
 - \frac{8}{9}\frac{\alpha_{Z^\prime}\alpha_s}{M_{Z^\prime}^2}
      \left[\frac{\hat{s}^2}{\hat{t}}+\frac{\hat{s}^2}{\hat{u}}\right]
      (b^2+{b^\prime}^2)\\
&& + \frac{\alpha_{Z^\prime}^2}{M_{Z^\prime}^4}\left[\frac{8}{3}
(b^4+{b^\prime}^4) \hat{s}^2 +
2b^2{b^\prime}^2(\hat{u}^2+\hat{t}^2)\right]\nonumber\\

&\frac{d\hat{\sigma}}{d\hat{t}}(qq^\prime\to qq^\prime) =&
 \frac{4\pi}{9\hat 
s^2}\alpha_s^2\left(\frac{\hat{s}^2+\hat{u}^2}{\hat{t}^2}\right)
 + \frac{\alpha_{Z^\prime}^2}{M_{Z^\prime}^4}
 \left[(b^4+{b^\prime}^4)\hat{s}^2+2b^2{b^\prime}^2\hat{u}^2\right]\nonumber

\end{array}
\end{equation}
\narrowtext
\noindent plus the cross sections related by crossing.
 Here, $q$ and $q^\prime$ denote quarks of different flavors;
$\alpha_{Z^\prime}=g_{Z^\prime}^2/4\pi$.

\begin{figure}
\centerline{\epsfig{file=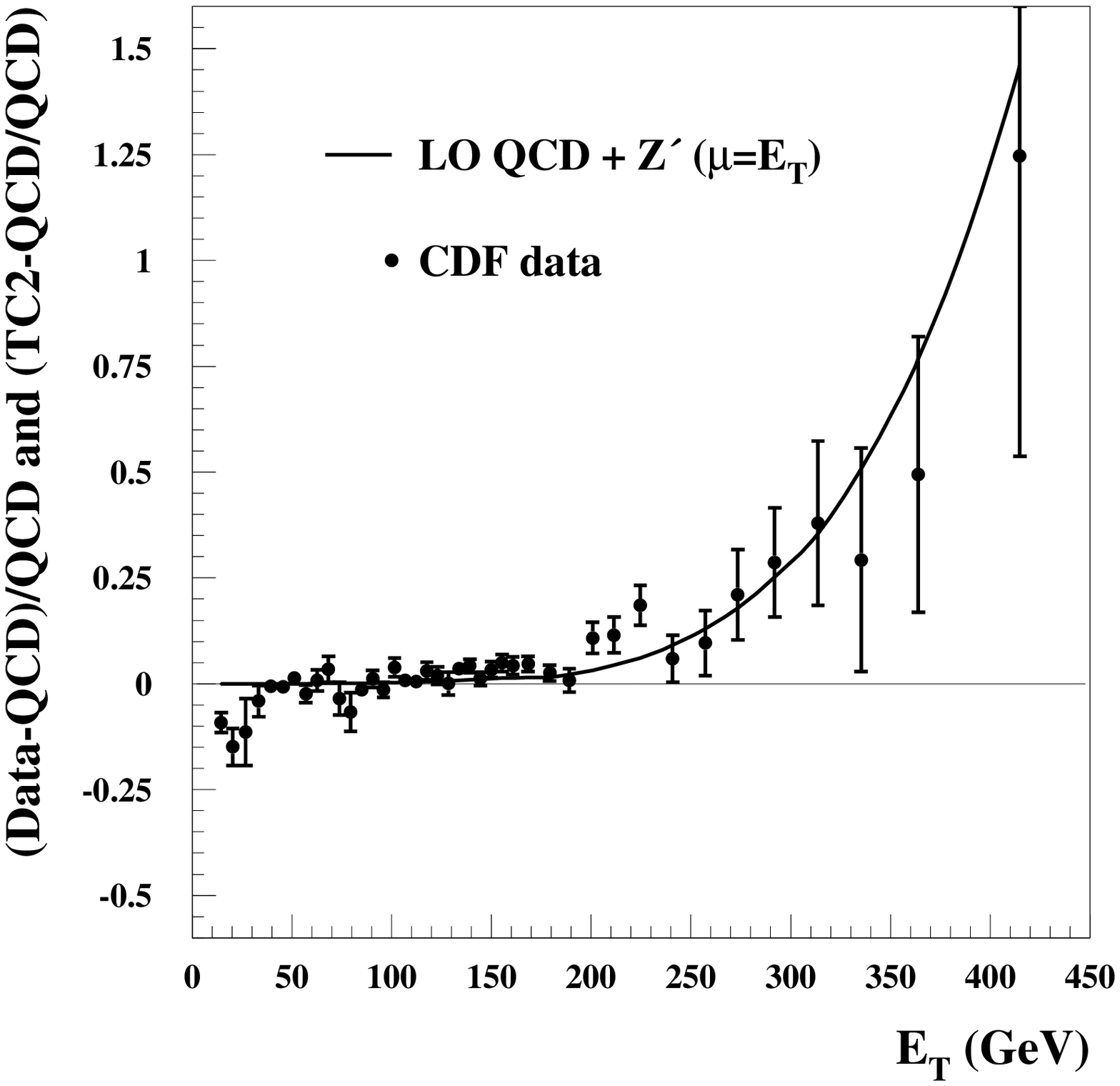,height=8cm,width=8cm}}
\caption{
  Difference plot of (data-QCD)/QCD and (TC2-QCD)/QCD for the inclusive 
jet cross section as a
  function of jet transverse energy $E_T$ in the central pseudorapidity
  region $0.1 < |\eta| < 0.7$. Points with statistical (only) error
bars are CDF data~\protect\cite{cdf:jetet}. The solid curve 
  shows the LO QCD plus the $Z^\prime$ gauge boson in TC2 models
  with $b^2+b'^2=4$ (best-fit value)
  and $\alpha_{Z'}/M^2_{Z'}=0.075\mbox{ TeV}^{-2}$.}
\label{fig1}
\end{figure}

\paragraph*{}
Using Eq.~(4), we computed the high-$E_T$ jet rate and compared it to
lowest-order QCD (LO~QCD) in Fig.\ref{fig1}. To compare with the CDF data,
also shown there with statistical errors only~\cite{cdf:jetet}, we
integrated over the jet pseudorapidity region $0.1 < |\eta| < 0.7$, and
used the MRSD0$^\prime$ parton distribution functions\protect\cite{MRS}
with the renormalization scale $\mu=E_T$. We varied the $U(1)_1$
hypercharges $b$ and $b'$ while fixing $\alpha_{Z'}/M^2_{Z'}=0.075\mbox{
TeV}^{-2}$. (This is not a restriction, since the cross section depends
only on products of $\alpha_{Z'}b^2/M^2_{Z'}$ and
$\alpha_{Z'}b'^2/M^2_{Z'}$. For $M_{Z'} = 2\mbox{ TeV}$, this corresponds
to the moderately strong coupling $\alpha_{Z'} = 0.3$.) We normalized our
results by fitting to CDF data in the region $40<E_T<120$~GeV where jet
production is dominated by $t$ and $u$-channel gluon exchange. For a
$\chi_{35}^2 \equiv \chi^2/\mbox{d.o.f.}=1.2$ ($\mbox{d.o.f.}=35$), we
found that the best-fit choice of parameters satisfies $b^2+b'^2=4$. 
There is a a 20\% probability of obtaining a larger value
of $\chi^2_{35}$. The region with 
$\chi^2_{35} < 3.0$ corresponds to $1.0<b^2+b'^2<5.8$. 
For comparison, the QCD prediction (the horizontal line
at zero in Fig.\ref{fig1}) has $\chi^2_{37} = 3.4$. Note that, because CDF
did not provide numerical values for systematic errors on their jet-$E_T$
data, we have not included them in determining $\chi^2_{35}$.

\paragraph*{}
A more stringent limit on $Z'$ parameters can be obtained by performing the
same analysis with the CTEQ4HJ set of parton distribution
functions\protect\cite{CTEQ}. These distributions were obtained after CDF
released its jet-$E_T$ data by including that in the overall fit. These
distribution functions have more gluons at high-$x$ than earlier sets. With
$\alpha_{Z'}/M^2_{Z'} =0.075\mbox{ TeV}^{-2}$ as before, the minimum
$\chi_{35}^2 \simeq 2$ occurs for $b^2+b'^2 < 1$. As above, systematic
errors were not included. We understand that this means the significance of
our fits to this data are not very meaningful. Rather, our point here is
that one can obtain a good fit to the CDF jet-$E_T$ spectrum with either a
TC2 $Z'$ or a set of distribution functions designed to fit the data. In
the rest of this paper, we are going to discuss results obtained with the
same MRS functions the CDF collaboration used in their published plots.

\begin{figure}
\centerline{\epsfig{file=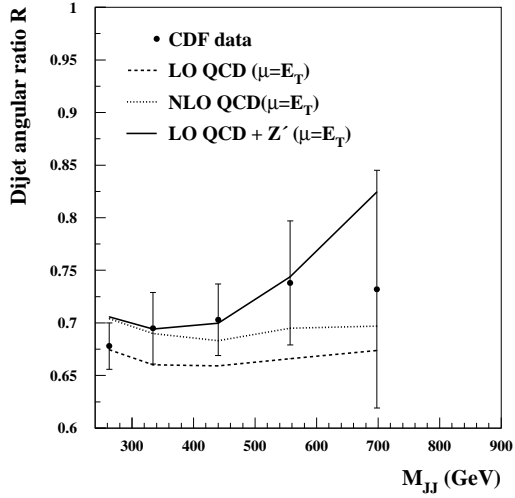,height=8cm,width=8cm}}
\caption{  Dijet angular ratio $R$
  as a function of the dijet invariant mass. Points with error bars are
CDF data~\protect\cite{cdf:jetang}; statistical and systematic errors are
added in quadrature. The solid curve shows the LO QCD plus the
  extra gauge boson $Z^\prime$ in a TC2 model with $b^2+b'^2=4$ and
$\alpha_{Z'}/M^2_{Z'}=0.075\mbox{ TeV}^{-2}$. The normalization of this
curve is described in the text.}
\label{fig3}
\end{figure}

\paragraph*{}
We also studied the dijet angular distribution dependence on the TC2
parameters. The CDF Collaboration has measured the 
ratio~\cite{cdf:jetang,David:Soper}
\begin{eqnarray}\label{ang:ratio}
R=\frac{\displaystyle \int^{0.46}_0d\eta^\star\frac{d^2\sigma}
                                          {dM_{jj}d\eta^\star}}
  {\displaystyle \int^{0.80}_{0.46}d\eta^\star\frac{d^2\sigma}
                                          {dM_{jj}d\eta^\star}} \ts\ts,
\end{eqnarray}
where $\eta^\star=(\eta_1-\eta_2)/2$ is the pseudorapidity of a jet in the
subprocess center-of-mass frame, $\eta_i$ is the $i$-th jet rapidity, and
$M_{jj}$ is the dijet invariant mass. This ratio is fairly insensitive to
the choice of parton distribution functions; we used MRSD0'. We have
computed the same quantity using the best-fit parameters determined above,
and the results are compared with the experimental data in Fig.\ref{fig3}.
In this figure, we followed the CDF procedure~\cite{cdf:jetang} of
normalizing the LO~QCD~$+$~$Z'$ curve by multiplying it, in each $M_{jj}$
bin, by the ratio of the NLO~QCD to LO~QCD $R$-values in that bin. The CDF
data points in this case include systematic errors, added in quadrature
with statistical ones. For the $Z'$ fit, we have $\chi^2_3 \approx 1.3$,
whereas the NLO~QCD fit has $\chi^2_5 \approx 0.69$. In terms of
probabilities, 30\% of fits will give $\chi^2_3 > 1.3$, while approximately
60\% will give $\chi^2_5 > 0.69$.

\begin{figure}
\centerline{\epsfig{file=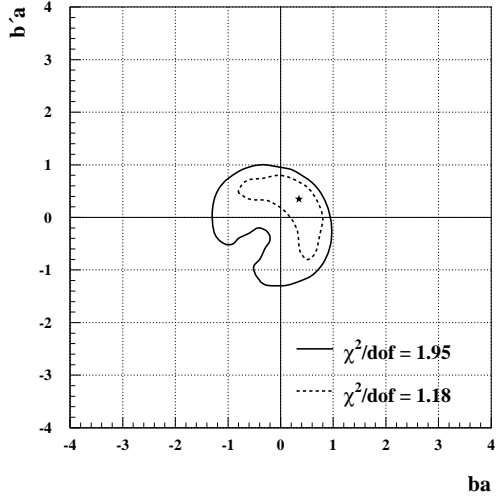,height=8cm,width=8cm}}
\caption{Parameter space regions allowed by CDF Drell-Yan data.
For $\alpha_{Z'}/M^2_{Z'}=0.075\mbox{ TeV}^{-2}$, 32\% of
fits have $\chi^2_4 > 1.18$ (dashed curve) while 10\% have $\chi^2_4 >
1.95$ (solid curve). The star corresponds to the best-fit and has $\chi^2_4
= 0.84$.}

\label{fig4}
\end{figure}

\begin{figure}
\centerline{\epsfig{file=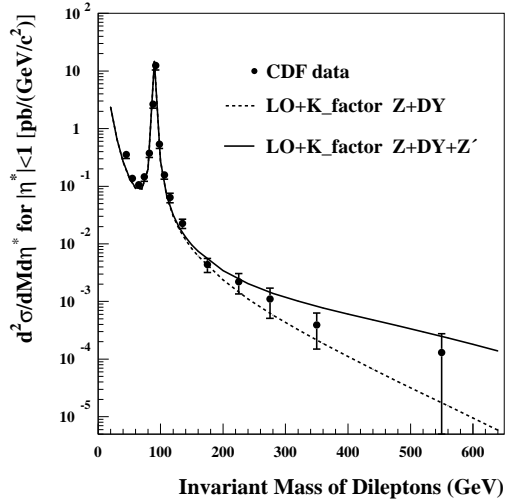,height=8cm,width=8cm}}
\caption{  Drell-Yan dilepton ($ee$, $\mu\mu$) pair production cross section 
$d^2\sigma/dMd\eta^*$ (averaged over $|\eta^*|<1$) as a 
function of the dilepton invariant mass.
Points with error bars are CDF data~\protect\cite{cdf:dy}.
The solid curve shows the LO QCD plus the extra gauge boson $Z^\prime$
with $\alpha_{Z'}/M^2_{Z'}=0.075\mbox{ TeV}^{-2}$, $ab=0.9$, $ab'=0$.
For comparison, the LO~QCD curve is shown by the dashed line.}

\label{fig5}
\end{figure}

\paragraph*{}
The CDF measurements of the Drell-Yan rate~\cite{cdf:dy} provide strong
constraints on the $Z'$ couplings to leptons. The subprocess cross section
is
\begin{equation}
\frac{d\hat{\sigma}}{d\hat{t}}(q_i\bar{q}_i\to \ell^-\ell^+)=
  \frac{\pi\alpha^2}{3\hat{s}^2}
   \left[{\cal A}_i(\hat{s})\left(\frac{\hat{u}}{\hat{s}}\right)^2
         +{\cal 
B}_i(\hat{s})\left(\frac{\hat{t}}{\hat{s}}\right)^2\right],
\end{equation}
where
\begin{eqnarray}
{\cal A}_i(\hat{s}) &=&
  \left[Q_i+\frac{4}{\sin^2{2\theta_W}}
   (T_{3i}-Q_i\sin^2{\theta_W})   
\left(\frac{1}{2}-\sin^2{\theta_W}\right)
\left(\frac{\hat{s}}{\hat{s}-M_Z^2}\right)
   + \frac{\hat{s}}{M_{Z^\prime}^2}
       \frac{\alpha_{Z'}}{\alpha}ba\right]^2 \nonumber\\
 & &
  +\left[Q_i+Q_i\tan^2{\theta_W}
        \left(\frac{\hat{s}}{\hat{s}-M_Z^2}\right)
  + \frac{\hat{s}}{M_{Z^\prime}^2}
       \frac{\alpha_{Z'}}{\alpha}
   b^\prime a\right]^2\\
{\cal B}_i(\hat{s}) &=&
  \left[Q_i-\frac{1}{\cos^2{\theta_W}}(T_{3i}-Q_i\sin^2{\theta_W})
   \left(\frac{\hat{s}}{\hat{s}-M_Z^2}\right)
  + \frac{\hat{s}}{M_{Z^\prime}^2}
     \frac{\alpha_{Z'}}{\alpha}ba\right]^2\nonumber\\
 & &
  + \left[Q_i-\frac{1}{\cos^2{\theta_W}}Q_i
     \left(\frac{1}{2}-\sin^2{\theta_W}\right)
      \left(\frac{\hat{s}}{\hat{s}-M_Z^2}\right)
     + \frac{\hat{s}}{M_{Z^\prime}^2}
        \frac{\alpha_{Z'}}{\alpha} b^\prime a\right]^2.
\end{eqnarray}
Following CDF's usage, we computed the LO~QCD-plus-$Z'$ Drell-Yan rate
using the MRS(A) parton distribution functions~\cite{MRS}. We obtained good
agreement with the CDF data below dilepton invariant mass $M = 120\mbox{
GeV}$ by multiplying the cross section by $K = 1.5$. (A simple,
multiplicative ``$K$-factor'' of this magnitude typically brings the LO~QCD
calculation of the Drell-Yan rate into agreement with the NLO calculation
and with low-energy data.) The CDF data was then used to constrain the
hypercharge products $ab$ and $ab'$. The allowed regions corresponding to
fits with $\chi^2_4 = 1.18$ (corresponding to a 32\% probability of
obtaining a worse fit) and $\chi^2_4 = 1.95$ (10\% probability) are shown
in Fig.\ref{fig4}. In Fig.\ref{fig5} we show the CDF data and the computed
cross section for a choice of parameters with $\chi^2_4 = 1.95$:
$\alpha_{Z'}/M^2_{Z'} = 0.075\mbox{ TeV}^{-2}$, $ab = 0.9$ and $ab' =0$.
For comparison, the LO QCD calculation with $K =1.5$ is also shown in the
figure; it has $\chi^2_6 = 0.94$ (i.e., approximately 50\% of fits will
have larger $\chi^2_6$).

\paragraph*{}

Although no fully acceptable TC2 model has been constructed so far (a
typical problem is the presence of unwanted Goldstone bosons), it is
interesting to apply our constraints to a set of parameters that guarantee
anomaly cancellation for the gauge interactions based on
Eq.~(1)\protect\cite{lane:japan}: $a=-1.6983$, $b=0.5157$, $b'=0.6233$.
Chivukula and Terning\protect\cite{Chiv_Tern} have found that consistency
of this model with precision electroweak measurements requires
$M_{Z'}>2.7\mbox{ TeV}$. We cannot set bounds on $\alpha_{Z'}$ and $M_{Z'}$
separately, but the inclusive-jet test provides a best-fit ($\chi_{35}^2 =
1.2$) value for this ratio which is comfortably within the limits we
obtained above in the sense that the $Z'$ mass constraint is less stringent
for the same value of $\alpha_{Z'}$: $\alpha_{Z'}/M_{Z'}^2 = 4/((0.52)^2 +
(0.62)^2) \times 0.075\mbox{ TeV}^{-2} = 0.46\mbox{ TeV}^{-2}$.

A much more stringent bound comes from the Drell-Yan data, which require
$\alpha_{Z'}/M_{Z'}^2< 0.021\mbox{ TeV}^{-2}$ for $\chi^2_4 = 1.95$. This
constraint cannot be satisfied unless either $M_{Z'} > 4\mbox{ TeV}$ or
$\alpha_{Z'} \ll 1$. In either case, there is unacceptable fine tuning
because the topcolor breaking scale is much larger than the top-quark mass
that it generates or the $SU(3)_1$ coupling must be set very close to its
critical value for chiral symmetry breaking~\cite{cdt}.  Although this
specific choice of parameters is ruled out, the same conclusion is not
necessarily true for the whole class of TC2 models. It remains an open
question whether a TC2 model exists that is both experimentally allowed and
theoretically palatable.

\thanks{}
We thank Kaori Maeshima and Robert Harris for much help and for guiding us
to data available from the CDF collaboration. We thank John Butler for
several useful conversations and Elizabeth Simmons for providing her
computer code. This research was supported in part by the Department of
Energy under Grant~No.~DE--FG02--91ER40676.

\appendix

\vfil\eject

\end{document}